\documentclass[preprint,aps,prb,showkeys,showpacs,draft]{revtex4}
\usepackage{graphics}
\usepackage{amsmath}
\begin{document}
\preprint{LA-UR-06-3959}
\title{Systematic quantum corrections to screening in thermonuclear fusion}
\author{Shirish M. Chitanvis}
\affiliation{
Theoretical Division,
 Los Alamos National Laboratory,
 Los Alamos, New Mexico 87545}
\date{\today}
\begin{abstract}
We develop a series expansion of the plasma screening length away from the classical limit in powers of $\hbar^{2}$. It is shown that the leading order quantum correction increases the screening length in solar conditions by approximately $2\%$ while it decreases the fusion rate by approximately $ 0.34\%$. We also calculate the next higher order quantum correction which turns out to be approximately $0.05\%$. 
\end{abstract}

\keywords{screening, fusion, plasma, quantum corrections}
\pacs{52.30.Ex, 52.27.Gr, 52.20.-j}

\maketitle

\newpage
\section{Introduction}
Salpeter\cite{salpeter54} wrote a seminal paper  more than half a century ago, concerning screening effects on thermonuclear reaction rates. He made the basic point that screening effects are small at the center of the sun.

There has been renewed interest more recently in utilizing the sun as a source of neutrinos to test the standard model of unification of electro-weak forces.  The measured neutrino flux deviates from predictions of the standard model by a factor of two\cite{bethe90}. The measurement uncertainty is\cite{bethe90,gruzinov98} $\sim 1\%$. Therefore it would be meaningful to quantify the theoretical estimate with equal precision. 
The structure and dynamics of the sun are complex\cite{bahcall94}. Various phenomena need to be identified and estimated correctly.  Screening of Coulomb repulsion between nuclei at extremely short distances is one of them. Many calculations of screening have been made since Salpeter's original paper, attempting to refine the degree of screening\cite{brown97,brown06,fiorentini04}, dynamic effects\cite{carraro88}, quantum fluctuations\cite{gruzinov98,gervino05}, etc\cite{bahcall02}.  Here we shall focus on quantum corrections to screening.  The most sophisticated calculation of this effect is that of Gruzinov and Bahcall\cite{gruzinov98}. 
In this paper, the electronic density matrix was evaluated accurately using Feynman's formulation in terms of a Schroedinger equation with the inverse temperature playing the role of imaginary time. Fermion statistics are ignored due to the high solar temperature. They sustain Salpeter's original conclusion that quantum corrections are minor.  These calculations are essentially correct, but cannot estimate in a systematic fashion the next higher order quantum correction.  We shall correct that deficiency in this paper.

This paper was written for the sake of completeness, since the super-Kamiokande experiment has been successful in obtaining evidence for neutrino mass (see Hosaka et al\cite{kam1} for recent results). 
Nevertheless, the results of this paper may yet prove useful for more precise quantitative interpretation of stellar experimental data\cite{pinsonn06}.

\section{Non-relativistic Quantum corrections}

We shall treat ions in a thermonuclear plasma as classical objects, while applying a quantum treatment to electrons.  The resulting partition function is evaluated as a deviation from the classical limit for electrons.  A series expansion is developed in powers of a dimensionless ratio involving $\hbar^{2}$.

Let us begin with the classical Poisson-Boltzmann equation for a single species of ions and electrons:

\begin{eqnarray}
 -\nabla^2 \phi &&= 4 \pi \rho \nonumber\\
\rho &&=\rho_{+}+\rho_{-}\nonumber\\
\rho_{+} &&=e ~ n~Z ~\exp(-Z e \phi/k_BT) \nonumber\\
\rho_{-}&&=- e~n~Z ~\exp(e \phi/k_BT) 
\label{pb1}
\end{eqnarray}
where $e$ is the magnitude of the electronic charge, $k_{B}$ is Boltzmann's constant, 
$n$ is the average number density,
$Z e$ is the ionic charge,
and $T$ is the temperature of the system.
We shall work in the linear regime, which is expected to apply in the solar interior\cite{brown06}, by retaining only terms first order in $\phi$:

\begin{eqnarray}
\nabla^2 \phi &&\approx \left(\frac{4 \pi n (Z^{2}n +Z n)e^{2}}{k_{B}T}\right) ~\phi\nonumber\\
&&\equiv \Lambda_{0}^{-2} \phi \nonumber\\
\Lambda_{0} &&= \sqrt {\frac{k_{B}T}{4 \pi n e^{2}(Z^{2}+Z)}}
\label{pb2}
\end{eqnarray}
where $\Lambda_{0}$ is the classical screening length.

The quantum-mechanical version of this approximate Poisson-Boltzmann equation for a single species of ions and electrons may be written in analogy with Eqn. \ref{pb1}:

\begin{eqnarray}
 -\nabla^2 \phi &&= 4 \pi \rho \nonumber\\
\rho &&=\rho_{+}+\rho_{-}\nonumber\\
\rho_{+} &&\approx  e ~ n~Z ~\left(1-\frac{Z e \phi}{k_{B}T}\right)\nonumber\\
\rho_{-}&&=- e~\vert \psi(\{\vec r\})\vert^{2}
\label{qpb1}
\end{eqnarray}
where $\psi$ is the many-body quantum wave-function for electrons, and $\{\vec r\}$ refers collectively to the electrons in the system, and $\phi$ is the electrostatic potential.

We now invoke the following scaled variables, in order to ease subsequent calculations:

\begin{eqnarray}
\tilde \phi &&= e \phi/k_{B}T \nonumber\\
\tilde \psi&&= \Lambda^{3/2}\psi\nonumber\\
\Lambda &&= \sqrt{\frac{k_B T}{4 \pi Z^{2}n e^2}}\nonumber\\
\vec r' && = \frac{\vec r}{\Lambda} \nonumber\\
\Gamma &&= \frac{e^{2}}{\Lambda k_{B}T}
\label{scl}
\end{eqnarray}
Note that the first of Eqns.\ref{scl} shows that we are using $k_{B}T$ as the energy scale.
The electrostatic potential is then given by:

\begin{equation}
 \nabla'^2 \tilde \phi = (\tilde \phi + 4 \pi \Gamma \vert\tilde \psi \vert^{2} - Z^{-1}) 
\label{sc2}
\end{equation}

This equation may be obtained from a Lagrangian density:

\begin{eqnarray}
{\cal L}_{0}&&= -\frac{1}{2} \vert \vec \nabla \tilde \phi\vert^{2}- v(\tilde \phi, \tilde \psi)\nonumber\\
v(\tilde \phi, \tilde \psi) &&= \frac{1}{2}\tilde \phi^{2} + 4 \pi \tilde \phi\Gamma \vert\tilde \psi \vert^{2} -Z^{-1}\tilde \phi
\label{sc3}
\end{eqnarray}

The corresponding Hamiltonian density can be easily derived:
\begin{equation}
{\cal H}_{0}= \frac{1}{2} \vert \vec \nabla \tilde \phi\vert^{2}+ v(\tilde \phi, \tilde \psi)
\label{sc3a}
\end{equation}

We will now introduce second-quantized notation to deal with the statistics of electrons:

\begin{equation}
v(\tilde \phi, \tilde \psi)  \to v(\tilde \phi, \tilde \psi_{\pm}) = \frac{1}{2}\tilde \phi^{2}- Z^{-1}\tilde \phi + 4 \pi \tilde \phi\Gamma (\tilde \psi^{\dagger}_{+}\tilde \psi_{+}+\tilde \psi^{\dagger}_{-}\tilde \psi_{-})
\label{sc3b}
\end{equation}
where $\tilde \psi_{\pm}$ are Grassmann variables, and the subscripts refer to the spin of the electrons.
The co-existence of Grassmann variables and scalars in Eqn.\ref{sc3b} is not problematic, since we shall use this discussion solely to define a partition function for the entire system. And soon thereafter we shall integrate over the electron degrees of freedom, so that only a functional involving the scalar potential survives.

The total Hamiltonian ${\cal H}$ for the system, including the quantum-mechanical part for the electrons is:

\begin{eqnarray}
{\cal H} = {\cal H}_{0} + {\cal H}_{Q} \nonumber\\
{\cal H}_{Q} &&= \Delta_{Q} (\vert \vec \nabla \tilde \psi_{+}\vert^{2}+\vert \vec \nabla \tilde \psi_{-}\vert^{2} )
\label{sc4}
\end{eqnarray}

The quantum correction has been encapsulated in the following dimensionless parameter:

\begin{equation}
\Delta_{Q}=\left( \frac{\hbar^{2}\Lambda^{-2}}{2 m k_{B}T}\right)
\end{equation}
where $m$ is the mass of the electron.

Since solar temperatures are $\sim {\cal O}(1 keV)$, and the rest energy of the electron is $0.55 MeV$, it follows that the non-relativistic approximation employed in Eqn.\ref{sc4} is valid.

The partition function may be written in scaled variables as:

\begin{equation}
{\cal Z} = \int~{\cal D}\tilde\phi~{\cal D}^{2}\tilde \psi_{\pm}~\exp(- \int d^{3}x' ({\cal H}_{0}+{\cal H}_{Q}))\label{pf1}
\end{equation}
where it is understood that $k_{B}T=1$ in the units we are using.

Note that the total number of electrons is associated with each ion of charge $Ze$ is $Z$, and is obtained via  $<(\tilde \psi_{+}^{\dagger}\tilde \psi_{+}+\tilde \psi_{-}^{\dagger}\tilde \psi_{-})>=Z n$ (where the angular brackets indicate an expectation value). We will indicate shortly how one may impose this constraint upon the system using a Lagrange multiplier. In condensed matter physics, this is done via the electronic chemical potential.  In our problem, it will turn out to be more convenient to institute this constraint via a functional involving just the electrostatic potential.

Note that the parameter $\Gamma$ is analogous to the usual plasma parameter.  It is much less than one for solar conditions.
For solar conditions, viz., a density of $100~g~cm^{-3}$, $T= 15 \times 10^{6}K$, $Z = 2$, it turns out that $\Lambda_{0} \approx 0.281 \AA$, $\Gamma \approx 0.04$. The value of $\Delta_{Q}$ turns out to be approximately $ 0.032$.  This is already an indication that quantum corrections will be small. In making these estimates, we have assumed a Helium plasma with thermodynamic properties similar to those at the center of the sun\cite{gruzinov98}.

The quadratic nature of the energy functional in Eqn.\ref{pf1} allows us to perform the functional integration over the Grassmann variables associated with the electronic degrees of freedom\cite{ramond}, allowing us to obtain:

\begin{eqnarray}
{\cal Z} &&\sim \int {\cal D}\tilde \phi ~\exp(-\int d^{3}x' ((1/2)\vert \vec \nabla \tilde \phi\vert^{2}+(1/2) \tilde\phi^{2} - Z^{-1}\tilde \phi))~ {\rm {Det}}({\cal F}) \nonumber\\
{\rm {Det}}({\cal F})&& = \exp (Tr \ln({\cal F}))\nonumber\\
{\cal F} && \equiv -\Delta_{Q}~\nabla^{2 } + 4 \pi \Gamma \tilde \phi
\label{eff1}
\end{eqnarray}

Having integrated over the electronic degrees of freedom, we are left with an effective energy density in terms of the electrostatic potential alone.  We choose to impose charge neutrality, which was discussed just below Eqn.\ref{pf1}, via a Lagrange multiplier by making the following addition ($\Delta {\cal H}$) to the energy density:

\begin{equation}
\Delta {\cal H} =  4 \pi \nu \tilde \phi
\label{lmul}
\end{equation}

Here $\nu$ may be interpreted physically as a uniform charge density, which will be adjusted to ensure the overall charge neutrality. 

We need to evaluate the determinant of the operator obtained in the process of of performing the quadratic functional integral over fermionic variables. This is conveniently performed in Fourier space:

\begin{equation}
Tr \ln({\cal F}) \equiv \int ~\frac{d^{3}k}{(2 \pi)^{3}}~ \ln (4 \pi \Gamma \hat \phi(k) + \Delta_{Q}~k^{2})
\label{eff2}
\end{equation}
where $\hat \phi(k)$ is the Fourier transform of $\tilde \phi$.

Now the estimates below Eqn.\ref{pf1} indicate that $4 \pi \Gamma >> \Delta_{Q}$ near the center of the sun,
so we propose a series expansion in powers of $\Delta_{Q}$:

\begin{equation}
\ln (4 \pi \Gamma \hat \phi(k) + \Delta_{Q}~k^{2}) \approx \ln(4 \pi \Gamma \hat \phi(k)) + \frac{\Delta_{Q}~k^{2}}{4 \pi \Gamma \hat \phi(k)}-\left( \frac{\Delta_{Q}~k^{2}}{4 \pi \Gamma \hat \phi(k)}\right)^{2}+... 
\label{q2}
\end{equation}

Furthermore, we will conform to the linear screening limit which is expected to apply in the solar interior\cite{brown06}.  By this we mean we shall seek an expansion of the determinant obtained above, to the quadratic order of the scalar potential.-- higher order terms can be accounted for using diagrammatic techniques.  We have the freedom to choose the value of the potential around which to perform the expansion.  We shall choose this value to be $\phi_{0}$, such that in the limit that $\hbar \to 0$, we recover the standard expression for the screening length obtained from the linearized Poisson-Boltzmann equation.  The power series to quadratic order yields:

\begin{eqnarray}
\ln (4 \pi \Gamma \hat \phi(k) + \Delta_{Q}~k^{2}) &&\approx \ln(4 \pi \Gamma) \nonumber\\
&&+( \ln (\phi_{0}) + (\hat \Phi(k)-\phi_{0})/\phi_{0} - (\hat \Phi(k)-\phi_{0})^{2}/(2 \phi_{0}^{2})) \nonumber\\
&&+ \frac{\Delta_{Q}~k^{2}}{4 \pi \Gamma \phi_{0}}~(1-(\hat \Phi(k)-\phi_{0})/\phi_{0}+ (\hat \Phi(k)-\phi_{0})^{2}/\phi_{0}^{2}) 
\label{q3}
\end{eqnarray}

The constant terms are unimportant as they can be absorbed into the normalization constant.
To be consistent, the rest of the effective energy density must also be expanded around $\phi_{0}$.  The net result would yield, in addition to a quadratic term, a term linear in $\hat \Phi$.  
The coefficient of this linear term is an effective background charge density, which must be zero in our neutral system. The coefficient can be set to zero by adjusting appropriately $\nu$, the Lagrange multiplier, defined in Eqn.\ref{lmul}.

With the proper charge neutrality constraint imposed, the screening length expression can be matched with the linearized classical Poisson-Boltzmann Eqn.\ref{pb2} when $\hbar \to 0$, by setting $\phi_{0} = \sqrt{Z}$.
Then the energy density may be written to leading order in $\Delta_{Q}$ in Fourier space as follows:

\begin{equation}
{\cal H} \approx \frac{1}{2}~k^{2}~(1 + \frac{2 \Delta_{Q}}{Z^{3/2}\Gamma}) \hat \Phi(k)^{2}+\frac{1}{2}(1 + 1/Z) \hat \Phi(k)^{2}
\label{qc3}
\end{equation} 

The corresponding Lagrangian density in real space is:

\begin{equation}
{\cal L} = -\frac{1}{2} ~(1 + \frac{2 \Delta_{Q}}{Z^{3/2}\Gamma}) \vert \vec \nabla \Phi(\vec r)\vert^{2} - \frac{1}{2}(1 + 1/Z) \Phi(\vec r)^{2}
\label{qc4}
\end{equation}

The equation of motion then follows:

\begin{eqnarray}
-\nabla^{2}\Phi + \ell^{-2}\Phi &&=0 \nonumber\\
\ell &&= \sqrt{\frac{1 + \frac{2 \Delta_{Q}}{Z^{3/2}\Gamma}}{ (1 + 1/Z)}}
\label{qc5}
\end{eqnarray}

In dimensionful units, the corrected screening length is:

\begin{equation}
\Lambda_{QC} = \sqrt{\frac{k_B T (1 + \frac{2 \Delta_{Q}}{Z^{3/2}\Gamma}) }{4 \pi (Z^{2} + Z) n e^2}}
\label{qc6}
\end{equation}

Note that without quantum corrections, the screening length is about $0.281 \AA$.
Quantum fluctuations increase the screening length by $\sim 2\%$ at solar conditions, defined earlier.
The classical enhancement factor of the fusion rate is $1.17$.
It is reduced slightly to $1.16$ via quantum corrections.
The next higher order quantum correction to the rate from our theory (expanding Eqn.\ref{q3} to ${\cal O}(\Delta_{Q}^{2})$) turns out to be approximately $0.05\%$. 

The numerical values obtained for the screening length and the Salpeter rate enhancement factor have been encapsulated for solar conditions in Table \ref{table1}:

\begin{table}[h]
\begin{tabular}{ll}
\hline
\textbf{Quantity} &  \textbf{Quantum Correction}\\
\hline
Screening Length & $ 2 \%$ \\
Rate enhancement & $0.34 \%$  \\
\hline
\end{tabular}
\caption{Comparison of classical and quantum-corrected quantities in solar conditions.}
\label{table1}
\end{table}
 
 Hence the leading order quantum correction decreases the fusion rate by about $0.34\%$ for the conditions chosen. The numerical estimates provided above are in fair agreement with those found in the literature\cite{salpeter54,gruzinov98}.
 
\section{Conclusion}

Systematic quantum corrections to screening in thermonuclear fusion were derived  in powers of $\hbar^{2}$, and estimated for solar conditions. Leading order corrections were shown to be less than $0.34\%$ under solar conditions, while the next leading order term is $\sim 0.05\%$.
Our corrections are consistent with those previously obtained by Gruzinov and Bahcall\cite{gruzinov98}.
They complement the results of Brown et al\cite{brown06} who show that classical, non-linear effects on screening are small.



\section{Acknowledgments}

I would like to thank A. Gruzinov for suggesting the problem, and C.J. ``Tex'' Tymczak for reading the manuscript critically.

This work was carried out under the auspices of the National Nuclear Security Administration of the U.S. Department of Energy at Los Alamos National Laboratory under Contract No. DE-AC52-06NA25396.

\bibliographystyle{apsrev.bst}

\bibliography{pre_scc_bibdata}
\newpage

\end{document}